%% file: emnlp2024.tex
\pdfoutput=1

\documentclass[11pt]{article}

\usepackage{acl}

\usepackage{times}
\usepackage{latexsym}

\usepackage[T1]{fontenc}

\usepackage[utf8]{inputenc}

\usepackage{microtype}

\usepackage{inconsolata}

\usepackage{graphicx}
\graphicspath{ {./images/} }

\usepackage{amsmath}

\usepackage{multirow}
\usepackage{booktabs}

\usepackage{makecell}
\usepackage{tabu}

\usepackage{amssymb} 

\usepackage{array}

\usepackage{multirow}
\title{Tool Graph Retriever: Exploring Dependency Graph-based Tool Retrieval for Large Language Models}

\author{Linfeng Gao$^{1}$, Yaoxiang Wang$^{1}$, Minlong Peng$^{2}$, Jialong Tang$^{3}$\\
\textbf{Yuzhe Shang$^{1}$, Mingming Sun$^{2}$, Jinsong Su$^{1}$}\\
  $^{1}$School of Informatics, Xiamen University, China, \\
  $^{2}$Baidu, Beijing, China, $^{3}$Alibaba, \\
  \texttt{\{gaolinfeng, shangyuzhe\}@stu.xmu.edu.cn}, \texttt{wyx7653@gmail.com}, \\
  \texttt{\{pengminlong, sunmingming01\}@baidu.com}, \\
  \texttt{tangjialong.tjl@alibaba-inc.com}, \texttt{jssu@xmu.edu.cn}
}

\begin{document}
\maketitle
\begin{abstract}

With the remarkable advancement of AI agents, the number of their equipped tools is increasing rapidly. However, integrating all tool information into the limited model context becomes impractical, highlighting the need for efficient tool retrieval methods.
In this regard, dominant methods primarily rely on semantic similarities between tool descriptions and user queries to retrieve relevant tools. However, they often consider each tool independently, overlooking dependencies between tools, which may lead to the omission of prerequisite tools for successful task execution.
To deal with this defect, in this paper, we propose \textbf{T}ool \textbf{G}raph \textbf{R}etriever (TGR), which exploits the dependencies among tools to learn better tool representations for retrieval. First, we construct a dataset termed TDI300K to train a discriminator for identifying tool dependencies. Then, we represent all candidate tools as a tool dependency graph and use graph convolution to integrate the dependencies into their representations. Finally, these updated tool representations are employed for online retrieval.
Experimental results on several commonly used datasets show that our TGR can bring a performance improvement to existing dominant methods, achieving SOTA performance. Moreover, in-depth analyses also verify the importance of tool dependencies and the effectiveness of our TGR.

\end{abstract}

\input{1-introduction}

\input{2-related_work}

\input{3-tool_graph_retriever}

\input{4-experiment}

\section{Conclusion}
In this paper, we introduce Tool Graph Retriever (TGR), leveraging tool dependencies to enhance the tool retrieval process for LLMs. We first define the criteria for tool dependency 
and establish a 
dataset to train a discriminator for identifying tool dependencies.
Then, 
we use this discriminator to handle candidate tools, forming a tool dependency graph.
Subsequently,
via graph convolution, we perform tool encoding based on this graph,
where the updated tool representations can be used for the final tool retrieval.
Experimental results and in-depth analyses strongly demonstrate the effectiveness of TGR across multiple datasets.

In the future, we will explore more features to improve our discriminator, which has a significant impact on the performance of our TGR.
Besides, we will try some efficient graph networks to obtain better tool representations.
Finally, how to further enhance the generalization of our TGR is also one of our future research focuses.

\section*{Limitations}
In our opinion, due to the absence of a tool dependency identification dataset, the accuracy of the discriminator is somewhat limited. The time complexity of graph construction is $O(N^2)$, which could be optimized by developing prior rules to filter out tools with no apparent dependency.

\bibliography{anthology, custom}

\appendix

\section{Evaluation on Different Similarity Computing Methods}

\begin{figure*}
    \centering
    \includegraphics[width=0.8\linewidth]{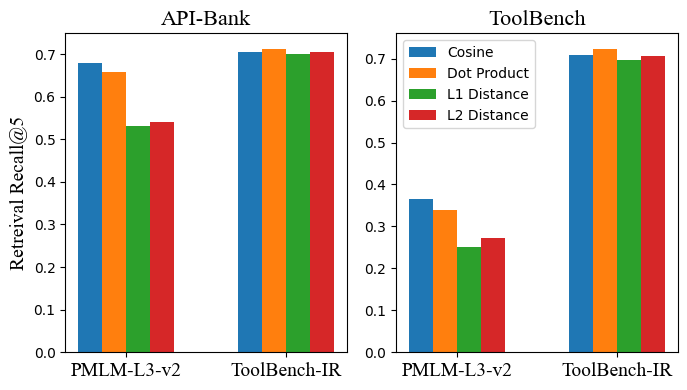}
    \caption{Evaluation on different similarity computing methods.}
    \label{fig:similarity}
\end{figure*}

We evaluate the effect of different similarity computing methods, including cosine similarity, dot product similarity, L1 distance, and L2 distance. The experiment is conducted on both API-Bank and ToolBench using Paraphrase-MiniLM-L3-v2 and ToolBench-IR as the text-embedding models. From Figure \ref{fig:similarity}, we can see that cosine similarity performs the best, while L1 and L2 distance have relatively lower performance. In our opinion, it is because the L1 and L2 distance ignore the angles between vectors, thus losing some features.

\section{Prompts for LLMs during Dataset Construction}
\label{sec:prompts}
\begin{figure*}
    \centering
    \includegraphics[width=0.8\linewidth]{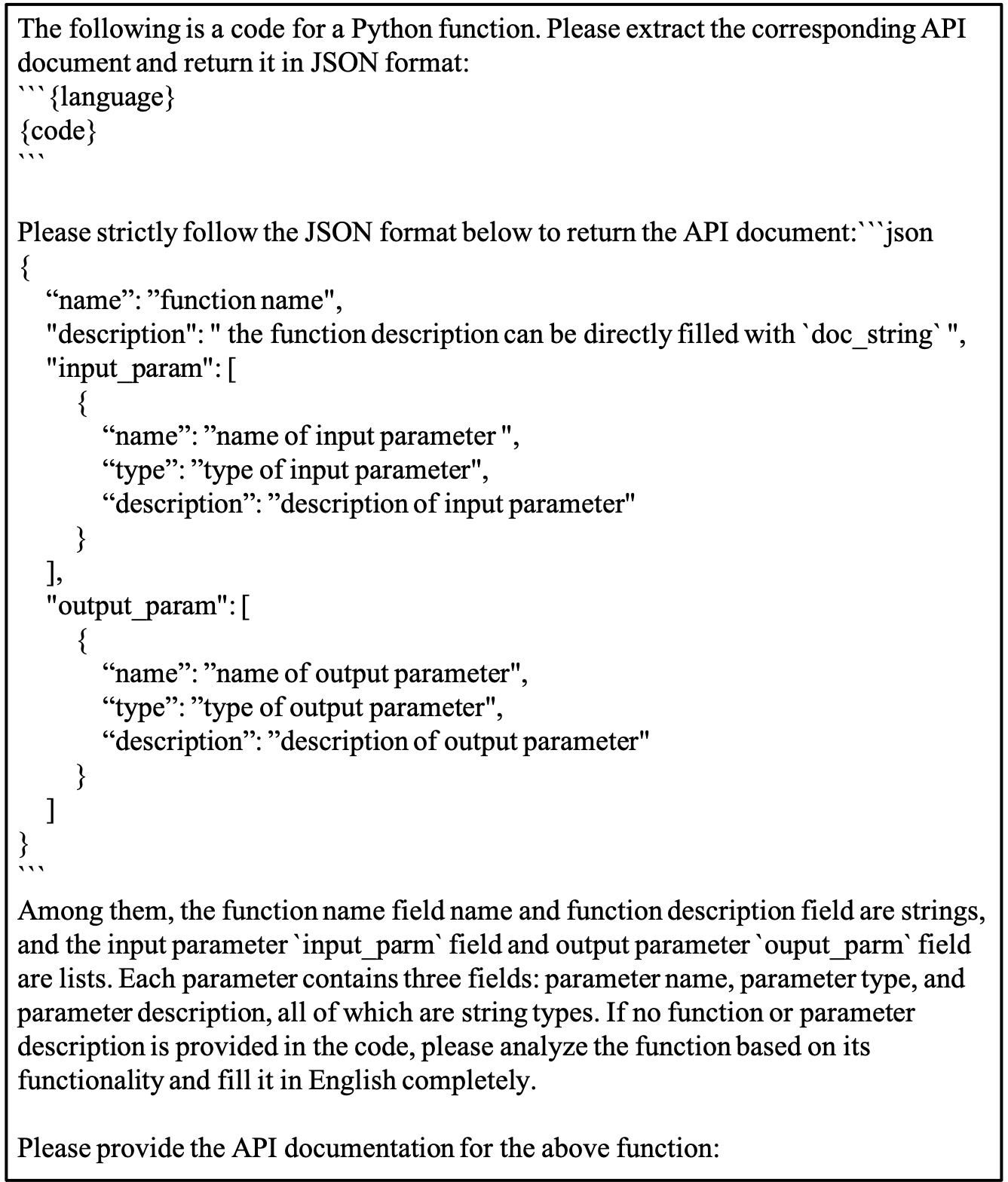}
    \caption{The prompt for the LLM to extract API documentation.}
    \label{fig:extract_prompt}
\end{figure*}

\begin{figure*}
    \centering
    \includegraphics[width=0.9\linewidth]{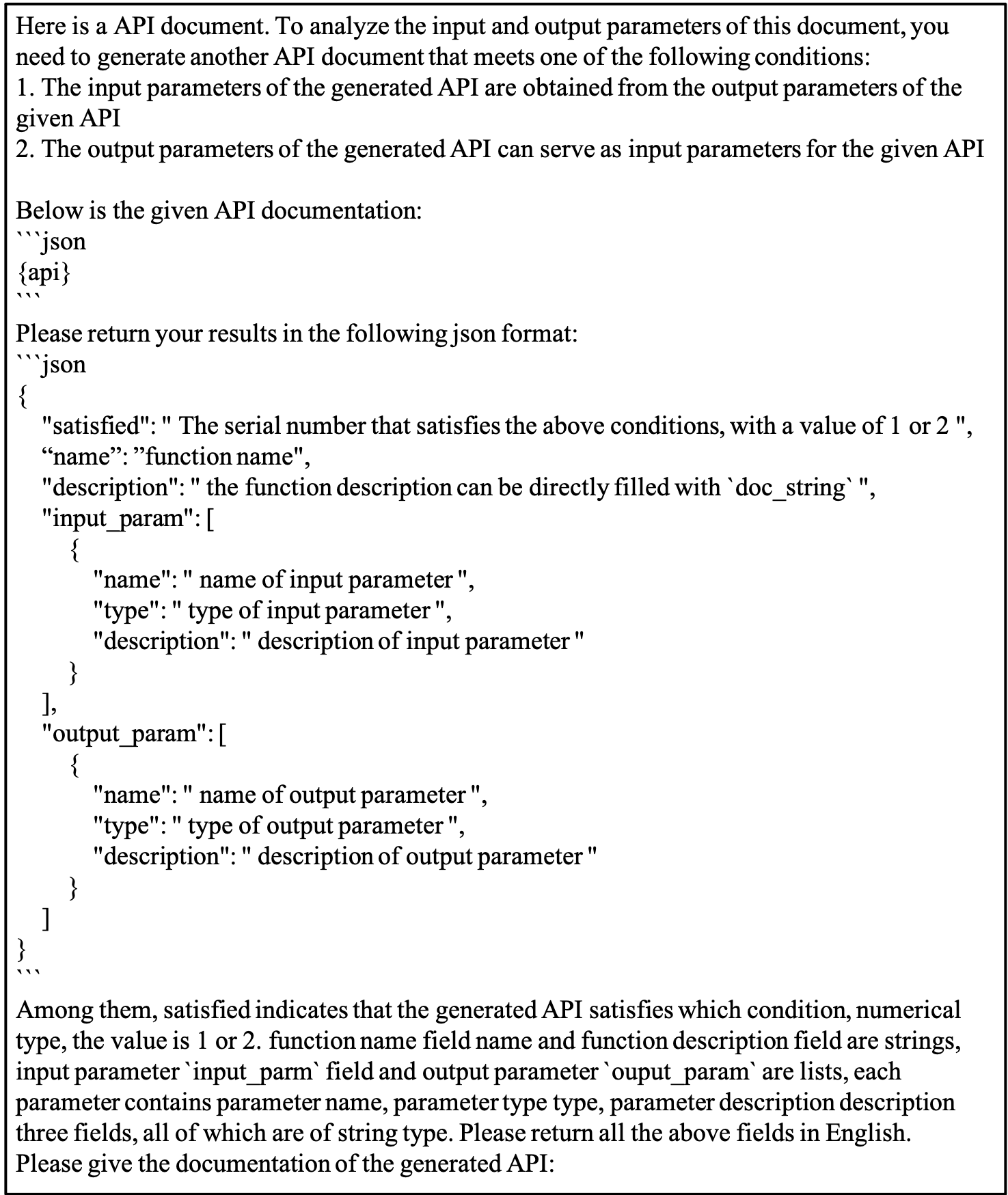}
    \caption{The prompt for the LLM to generate API documentation of a dependent tool function.}
    \label{fig:generate_prompt}
\end{figure*}

\begin{figure*}
    \centering
    \includegraphics[width=0.7\linewidth]{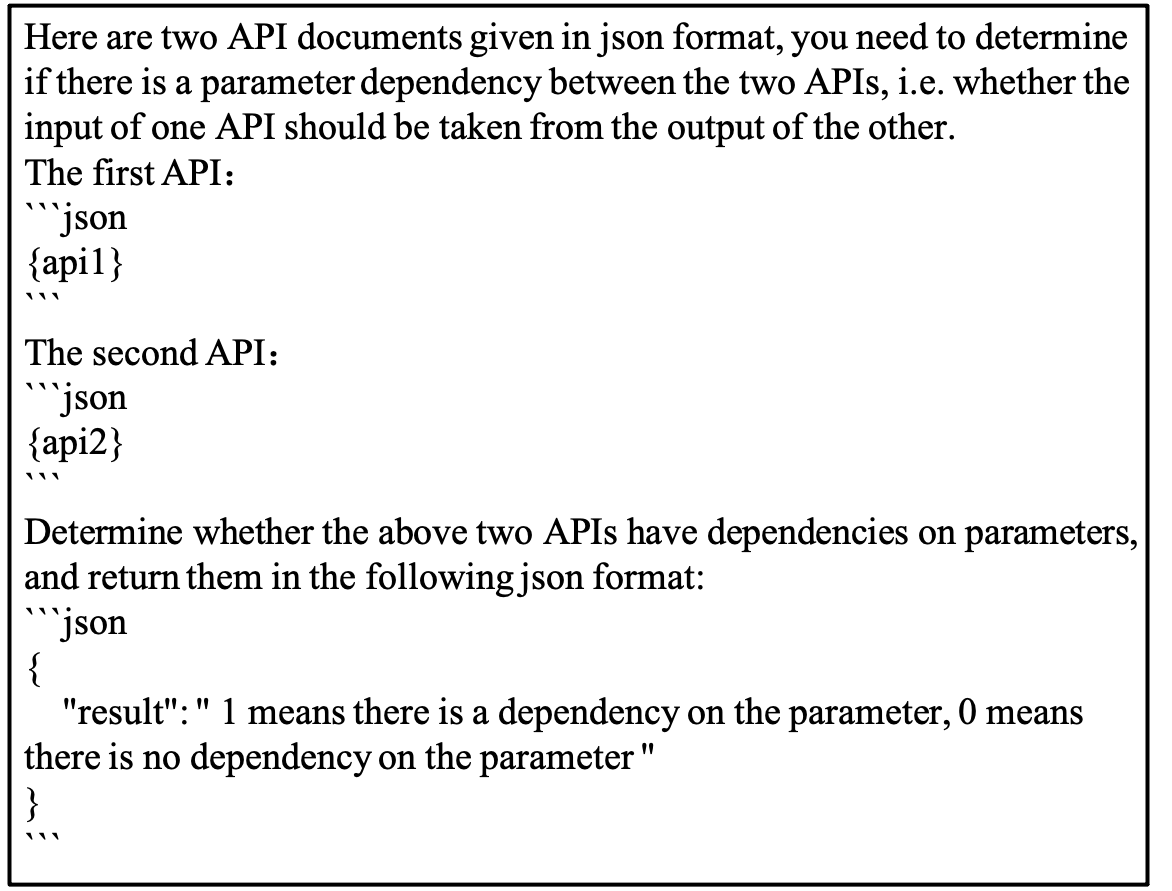}
    \caption{The prompt for the LLM to verify the dependency between two tools with the format of API documentation.}
    \label{fig:verify_prompt}
\end{figure*}

Here we provide the prompts we used for each LLM in the dataset construction pipeline we mentioned in Section \ref{sec:pretrain}. The prompts are shown in Figure \ref{fig:extract_prompt}, \ref{fig:generate_prompt}, and \ref{fig:verify_prompt} respectively.

\section{Visualization of the Graph}
\label{sec:vis}

\begin{figure*}[tbp]
    \centering
    \includegraphics[width=\linewidth]{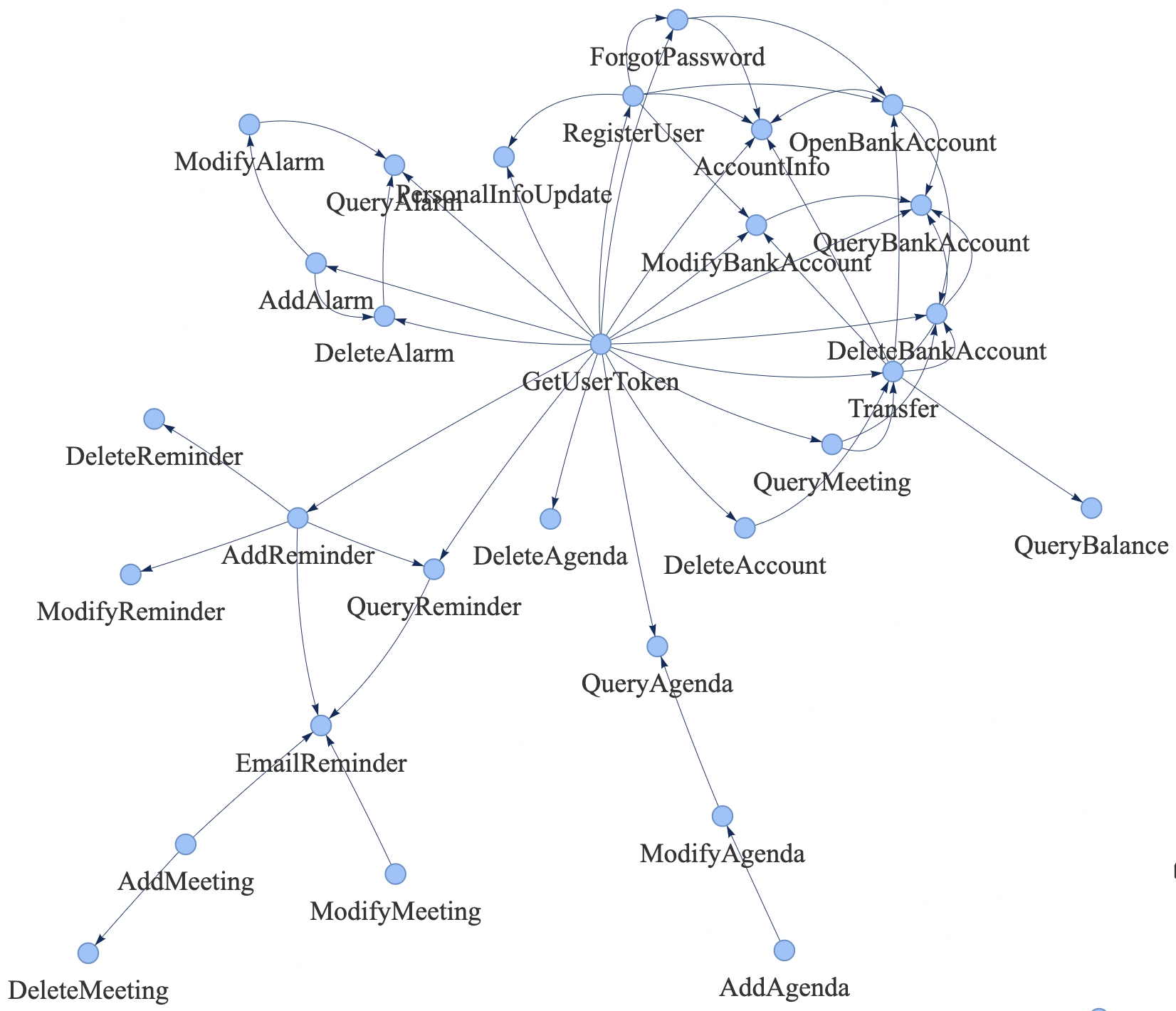}
    \caption{Visualization of the part of our constructed tool dependency graph in API-Bank \cite{li2023api}. The directed edge from $t_a$ to $t_b$ means $t_a$ is the prerequisite of $t_b$, i.e. the calling of $t_b$ depends on $t_a$.}
    \label{fig:graph_apibank}
\end{figure*}

\begin{figure*}[tbp]
    \centering
    \includegraphics[width=0.7\linewidth]{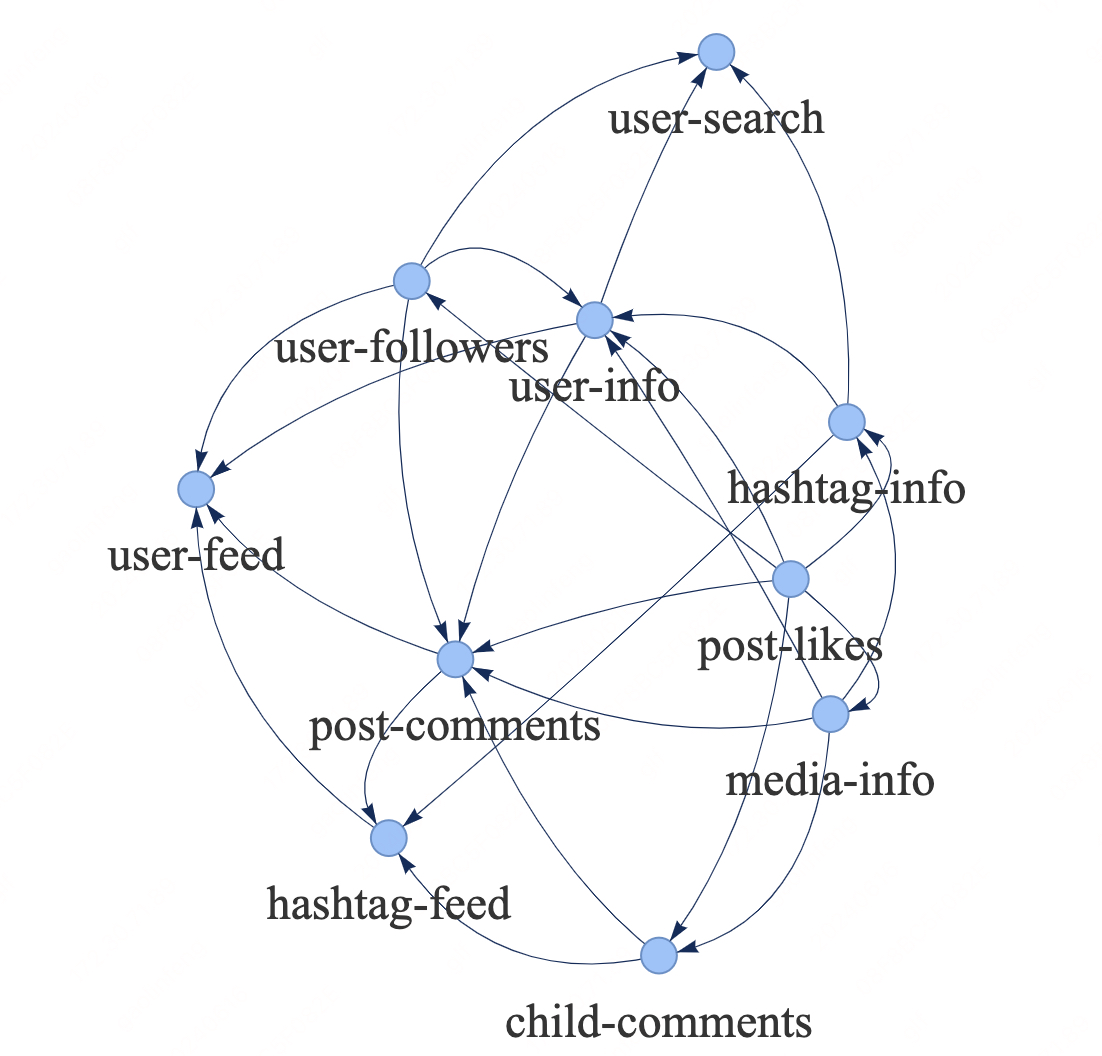}
    \caption{Visualization of the part of our constructed tool dependency graph in Toolbench \cite{qin2023toolllm}.}
    \label{fig:graph_toolbench}
\end{figure*}

We also visualize the tool dependency graph. Considering aesthetics and simplicity, we display part of the graph in API-Bank and ToolBench constructed by the discriminator. The graph is shown in Figure~\ref{fig:graph_apibank} and \ref{fig:graph_toolbench}.

\end{document}

%% file: 1-introduction.tex
\section{Introduction}
\label{sec:intro}

As an important step towards artificial general intelligence (AGI), tool learning expands the ability of LLM-based AI agents and enables them to interact with the external environment. \cite{goertzel2014artificial, dou2023towards, mclean2023risks}. However, as the number of equipped tools increases rapidly, it has become challenging for LLMs to process all the tool information, primarily due to the context length limitations. 
Therefore, a typical framework of AI agents employs a retriever to retrieve the candidate tools before the practical task, 
which involves the following four steps.
First of all, relevant tools are retrieved from the equipped tool set according to the task description provided by user. Secondly, the LLM, guided by a delicately-designed prompt and the tool retrieval results, creates a tool-invoking plan as the solution path for the task. Thirdly, it takes actions to invoke tools based on the plan and receives feedback from the tool execution result. Finally, if the task is considered complete, it will generate the final response to the user.

\begin{figure}
    \centering
    \includegraphics[width=\linewidth]{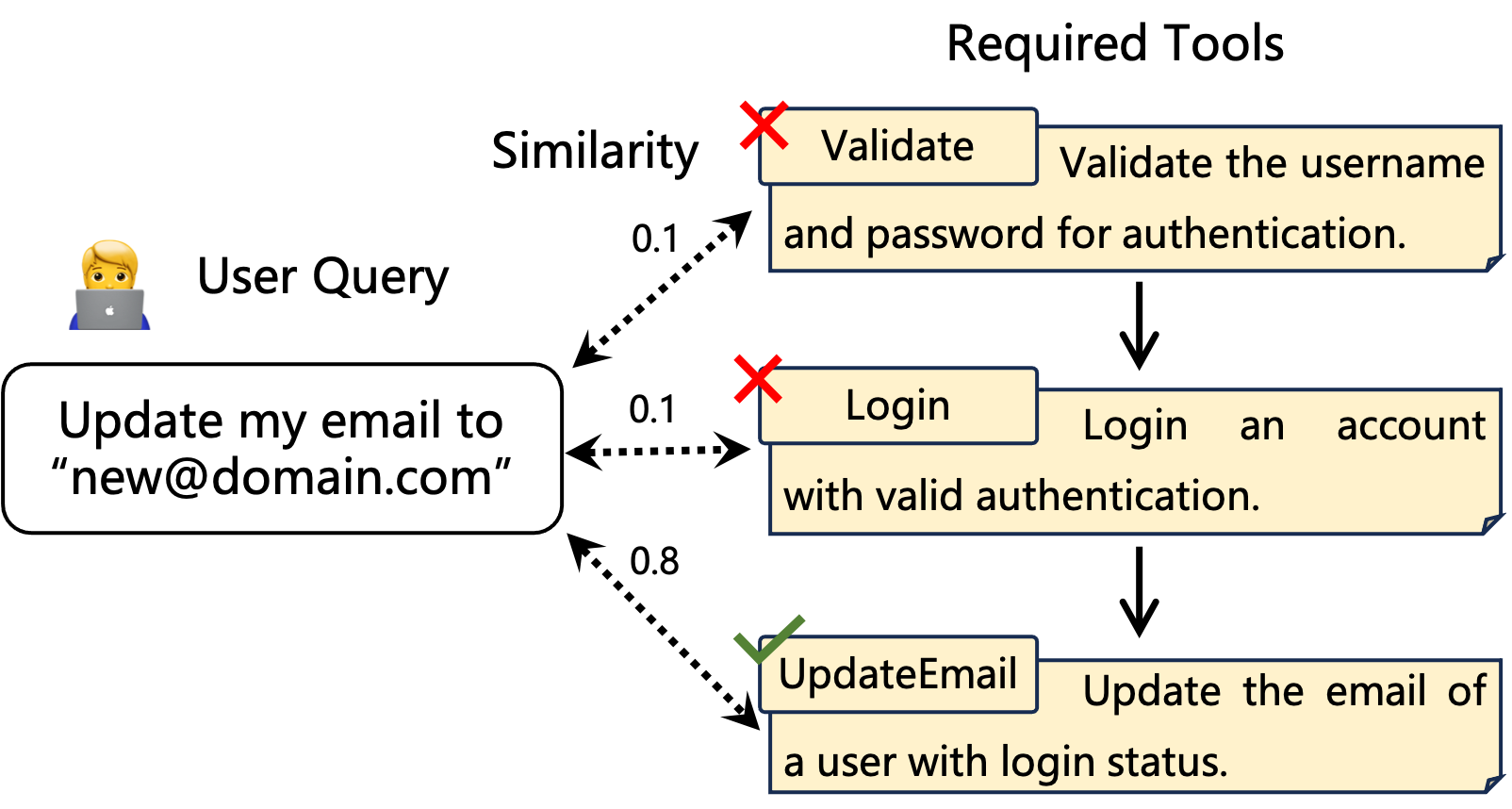}
    \caption{An example of dominant tool retrieval process, where some necessary prerequisite tools are omitted due to low semantic similarities. The down arrows $\downarrow$ denote the calling order of the tools.}
    \label{fig:example}
\end{figure}

\begin{figure*}[tbp]
    \centering
    \includegraphics[width=0.8\linewidth]{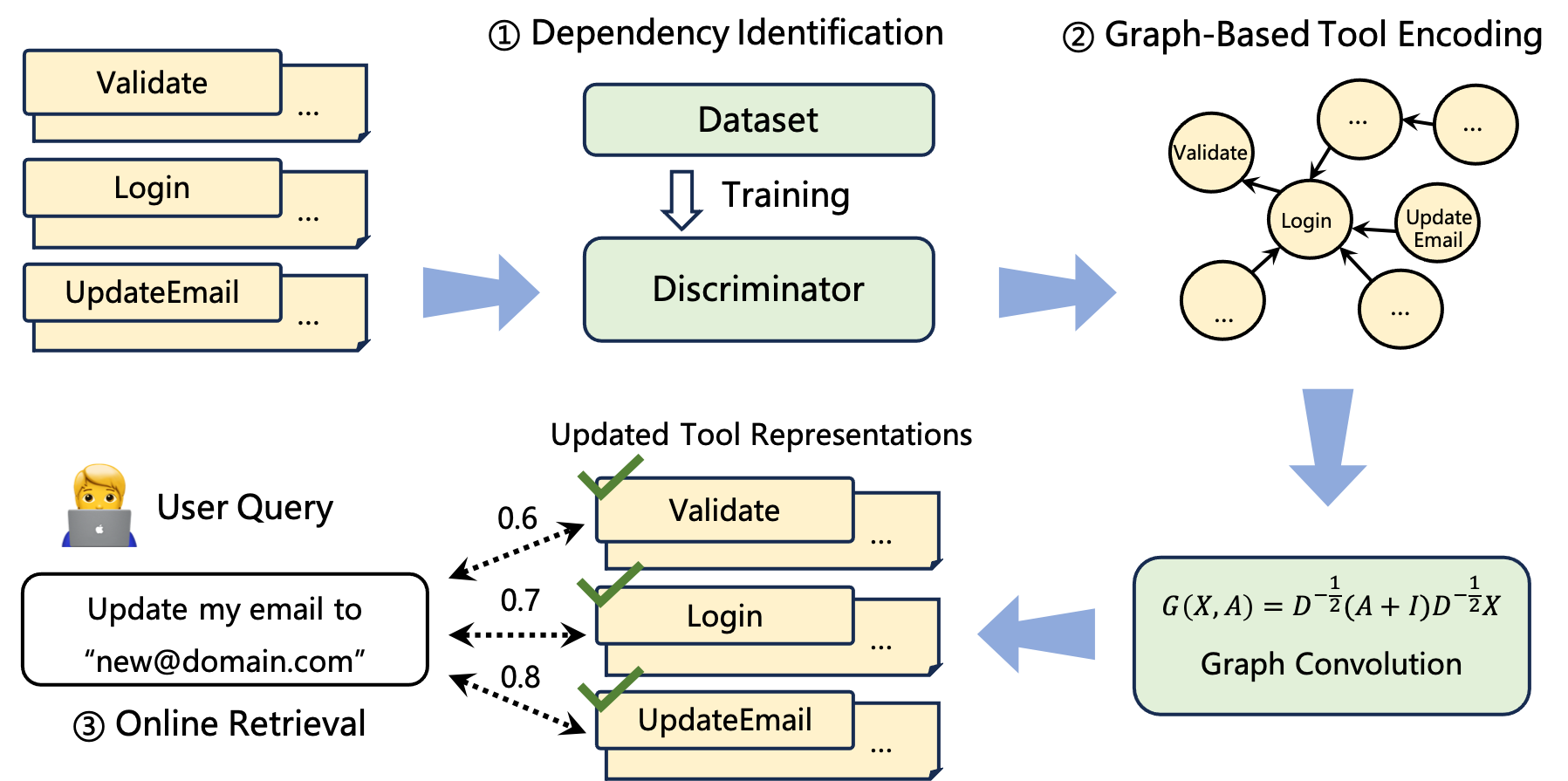}
    \caption{Our proposed TGR involves three steps: (1) Dependency Identification, where we build a dataset for tool dependency identification and train a discriminator; (2) Graph-Based Tool Encoding, where we represent the tools with dependencies as a graph and integrate the dependencies into tool representations with graph convolution; (3) Online Retrieval, where we utilize the updated tool embeddings to compute query-tool similarities as the final retrieval scores.}
    \label{fig:method}
\end{figure*}

As the first step in the above process, tool retrieval plays a critical role in constructing a high-performing tool-augmented agent. This is because the context length of the model restricts us to using only a limited number of tools. If necessary tools cannot be accurately retrieved, it will result in an execution error.
To achieve accurate tool retrieval, prevalent tool retrieval methods primarily focus on the semantic similarities between the tool descriptions and the user queries \cite{patil2023gorilla, li2023api, qin2023toolllm}. They consider each tool independently, which, however, results in the omission of some necessary prerequisite tools during retrieval. For instance, in the example of Figure \ref{fig:example}, the solution path for the query ``\textit{Update my email to `new@domain.com'.}'' involves three tools that should be invoked in sequence: ``\textit{Validate}'', ``\textit{Login}'', and ``\textit{UpdateInfo}''. However, the descriptions of tools ``\textit{Validate}'' and ``\textit{Login}'', which are about ``\textit{Validate credential}'' and ``\textit{Login account}'', are semantically irrelevant to the query. As a result, although the invocation of ``\textit{UpdateInfo}'' depends on the results of ``\textit{Validate}'' and ``\textit{Login}'', only the tool ``\textit{UpdateInfo}'' can be successfully retrieved.

In this paper, we propose \textbf{T}ool \textbf{G}raph \textbf{R}etriever (TGR), which exploits the dependencies between tools to refine the tool retrieval process. As shown in Figure \ref{fig:method}, it involves three steps: (1) Dependency Identification. In this step, we construct a dataset, termed as TDI300K, and train a discriminator to identify the tool dependencies; (2) Graph-Based Tool Encoding. To model the dependencies, we construct a graph with tools as nodes and their dependencies as edges. Then we use graph convolution to integrate the dependencies for a better learning of the tool representations; (3) Online Retrieval. We conduct online retrieval by calculating the query-tool similarity with the updated tool representations. Compared with previous studies \cite{li2023api, qin2023toolllm, patil2023gorilla}, TGR leverages the tool dependencies as additional information to refine the retrieval process, thus leading to better results.

Overall, our contributions can be summarized as follows:
\begin{itemize}
    \item We propose \textbf{T}ool \textbf{G}raph \textbf{R}etriever (TGR), leveraging tool dependencies as additional information to improve the performance of tool retrieval. 
    \item We construct a tool dependency identification dataset termed TDI300K and subsequently train a discriminator, facilitating the subsequent studies in this area.
    \item Experimental results and in-depth analyses on several commonly-used datasets demonstrate that TGR brings the improvement of Recall, NDCG and Pass Rate to existing dominant methods, achieving state-of-the-art performance on several commonly-used datasets.
\end{itemize}

%% file: 2-related_work.tex
\section{Related Work}

Recently, LLMs have demonstrated outstanding abilities in many tasks. Meanwhile, it becomes dominant to equip LLMs with external tools, deriving many tool-augmented LLMs such as Toolformer \cite{schick2023toolformer}, ART \cite{paranjape2023art} and ToolkenGPT \cite{hao2023toolkengpt}. However, as the number of tools grows rapidly, how to efficiently conduct tool retrieval becomes more important. 

In this regard, \citet{qin2023toolllm} employ Sentence-BERT \cite{reimers2019sentence} to train a dense retriever based on a pretrained BERT-base \cite{devlin2018bert}. The retriever encodes the queries and tool descriptions into embeddings respectively and selects top-$k$ tools with the highest query-tool similarities.
Similarly, \citet{li2023api} and \citet{patil2023gorilla} employ paraphrase-MiniLM-L3-v2 \cite{reimers2019sentence} and text-embedding-ada-002 \footnote{https://platform.openai.com/docs/guides/embeddings} as the tool retrievers respectively. 
Besides, \citet{hao2023toolkengpt} represent tools as additional tokens and finetune the original LLM to autonomously select the tool to be invoked.
Unlike the studies mentioned above, \citet{liang2023taskmatrix} divide tools into different categories to quickly locate relevant ones. They also employ Reinforcement Learning from Human Feedback for the entire task execution, so as to enhance the ability of the tool retriever.

Different from the above studies, TGR improves the effectiveness of tool retrieval with tool dependencies as additional information. We first identify the dependencies between tools and model them as a graph. Then, we use graph convolution to integrate the dependencies into tool representations, which are used for final online retrieval.
To the best of our knowledge, our work is the first attempt to leverage tool dependencies to refine the retrieval process.

%% file: 3-tool_graph_retriever.tex
\section{Tool Graph Retriever}

As shown in Figure \ref{fig:method}, the construction and utilization of our retriever involve three steps: 1) Dependency Identification; 2) Graph-Based Tool Encoding; 3) Online Retrieval. he following sections provide detailed descriptions of these steps.

\subsection{Dependency Identification}
\label{sec:discriminator}
In this work, we consider the tool $t_a$ depends on the tool $t_b$ if they satisfy one of the following conditions:
\begin{itemize}
    \item The tool $t_a$ requires the result from the tool $t_b$ as the input. For example, if we want to update the email of a user with the tool ``\emph{UpdateEmail}'', we should first get the permission from the user with the tool ``\emph{Login}''. Therefore ``\emph{UpdateEmail}'' depends on ``\emph{Login}'' for permission acquisition.
    \item The tool $t_a$ requires the tool $t_b$ for prior verification. For example, the tool ``\emph{Login}'' depends on the tool ``\emph{Validate}'' to ensure a valid username combined with the correct password.
\end{itemize}

Based on the definition above, we build a dataset termed TDI300K for tool dependency identification with the format $\{\langle t_a, t_b \rangle, y\}$, where $\langle t_a, t_b \rangle$ denotes a pair of tools and $y$ denotes their dependencies with three categories: (1) $t_a$ depends on $t_b$, (2) no dependency exists between $t_a$ and $t_b$ and (3) $t_b$ depends on $t_a$. It is worth noting that the dependencies between tools are sparse in the tool set, which, however, poses challenges for training the discriminator on a dataset with an imbalanced proportion of different categories. To solve this problem, we adopt a two-stage strategy to train a 3-class discriminator: the pretraining stage enables the discriminator to understand tool functions, and the finetuning stage enhances its ability to identify tool dependencies.

\begin{figure}
    \centering
    \includegraphics[width=\linewidth]{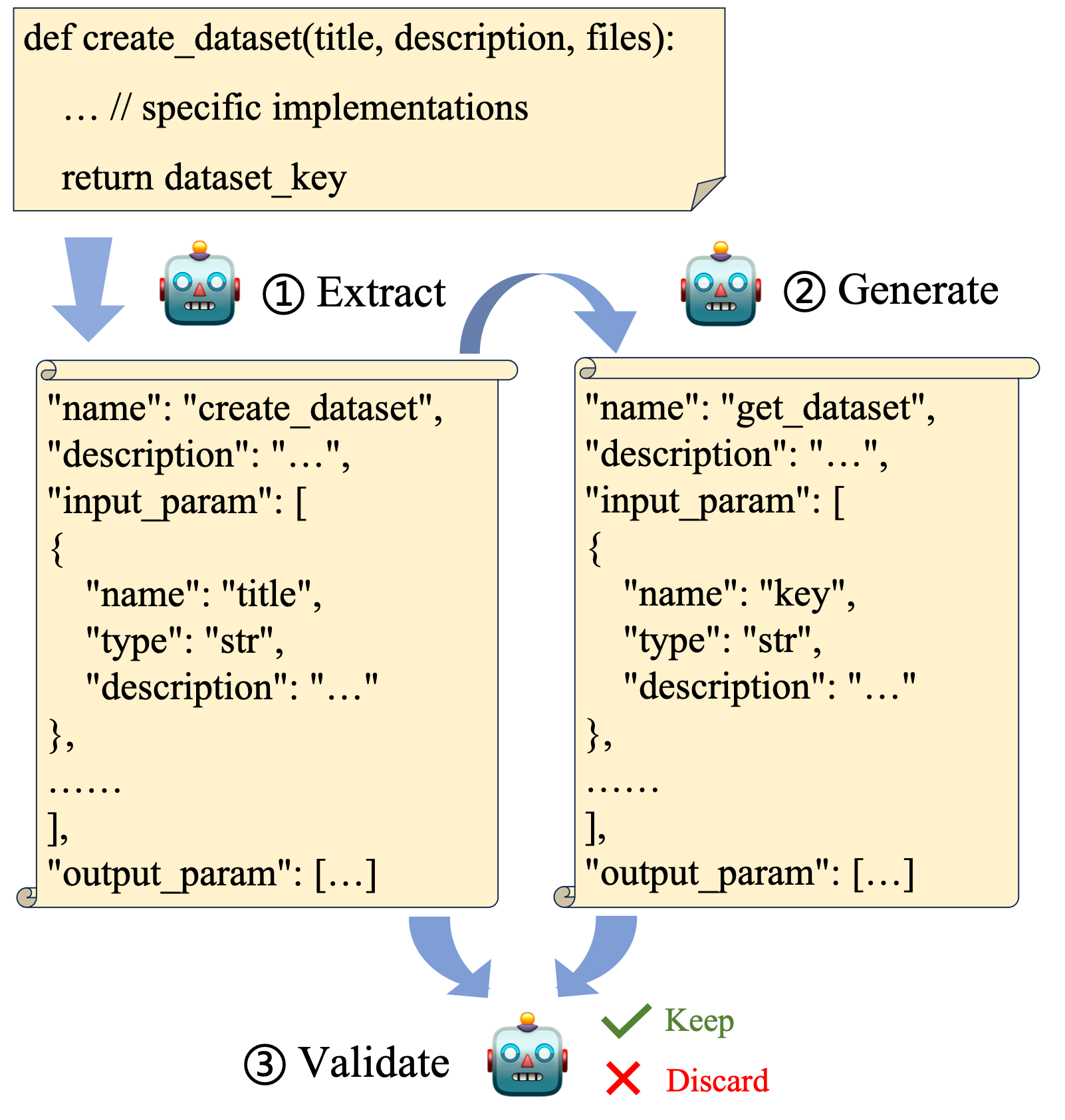}
    \caption{The pipeline used to construct the pretraining dataset, which involves three steps: 1) Extract tool document; 2) Generate dependent tool document; 3) Validate and filter the dependency.}
    \label{fig:framework}
\end{figure}

\paragraph{Pretraining}
\label{sec:pretrain}
Due to the lack of open-source tool dependency identification dataset, we design a three-step pipeline to construct the pretraining dataset derived from CodeSearchNet \cite{husain2019codesearchnet}, which contains 1.78 million real function implementations across various programming languages. As shown in Figure \ref{fig:framework}, we employ three agents based on gpt-3.5-turbo to extract tool documents, generate dependent tool documents, and validate the dependency. The LLM-specific prompts are shown in the Appendix \ref{sec:prompts}.
Firstly, given the specific implementation of a tool function, which is the source of $t_a$, we extract the document in JSON format, containing descriptions of tool functions, input parameters, and output results. Subsequently, the document of another tool $t_b$ is generated which is required to depend on $t_a$. Finally, we evaluate whether the dependency between $t_a$ and $t_b$ fulfills the predefined criteria, discarding tool pairs that do not satisfy the conditions.

Once we obtain an instance where $t_b$ depends on $t_a$, their positions can be swapped to obtain the opposing dependency category. Finally, we construct the instances without tool dependencies by breaking up and shuffling the tool pairs to make $t_a$ and $t_b$ independent. The statistics of the pretraining part of TDI300K are shown in Table \ref{tab:data}. Notice that here we keep three categories balanced to ensure a comprehensive learning of our discriminator on all three categories.

The pretraining process is a 3-class classification task, where we concatenate the documents of $t_a$ and $t_b$ and separate them with a special token {\tt [SEP]}, following \citet{devlin2018bert}. Besides, we add a special classification token {\tt [CLS]} before the input sequence, whose final hidden state is used for the classification task. With $\hat{y}$ denoting the model prediction, we define the following cross-entropy training objective:
\begin{equation}
    L(y,\hat{y})=-\sum_{k=1}^3y_k\log(\hat{y}_k).
\end{equation}

\begin{table}
\centering
\begin{tabular}{c|c|c}
\toprule
Category & Pretraining & Finetuning \\
\midrule
$t_a \rightarrow t_b$ & 92,000 & 1,029 \\
$t_a \times t_b $ & 92,000 & 33,365 \\
$t_a \leftarrow t_b$ & 92,000 & 1,056 \\
\bottomrule
\end{tabular}
\caption{The statistics of our constructed dataset TDI300K for tool dependency identification. The arrow $\rightarrow$ indicates the direction of the dependency and $\times$ means no dependency.}
\label{tab:data}
\end{table}

\paragraph{Finetuning}
To enhance the ability of the discriminator, we further finetune it on a manually constructed dataset with imbalanced category proportions which is more consistent with the real application scenario. First, we collect real function tools from open-source datasets, projects, and libraries\footnote{The sources include datasets like the training dataset of ToolBench, projects like online shopping, and libraries like OpenGL.}. Then, we write documents for these function tools with the same format as those in the pretraining dataset. Subsequently, these tools are organized as several tool sets to facilitate dependency annotation. Based on the definition of tool dependency mentioned above, we manually annotate the dependency categories given a pair of tools within a tool set. The statistics of the result datasets are also shown in Table \ref{tab:data}.

From Table \ref{tab:data}, we can clearly find that the imbalance category proportions propose a challenge for the discriminator. To deal with this problem and avoid overfitting, we define the following category-specific average training loss:
\begin{equation}
    L(y,\hat{y})=-\sum_{k=1}^3\frac{\sum_{i=1}^{N_k}y_{i,k}\log(\hat{y}_{i,k})}{N_k}
\end{equation}
where $N_k$ denotes number of instances with the $k$-th dependency category.

Notably, during the practical finetuning process, we split 20 percent of the whole dataset as the validation dataset, which is used to keep the checkpoint with the best performance. We also manually construct the testing dataset, which is derived from the existing tools in API-Bank, and the dependency categories are manually annotated. It contains 60, 500, and 60 samples for each category respectively. Here we choose API-Bank as the source of the test dataset since the tools are massive and the dependencies are hard to annotate in ToolBench. The performances of the discriminator on the validation and testing dataset will be presented in Section \ref{sec:exp_discriminator}.

\subsection{Graph-Based Tool Encoding}
With the above tool dependency discriminator, we use it to identify the dependencies among the tool set and then construct a tool dependency graph.
Formally, our graph is directed and can be formalized as $G=(V, E)$.
In the node set $V$, each node represents a candidate tool. As for the edge set $E$, if the tool $t_a$ depends on the tool $t_b$, the node of $t_a$ will be linked to that of $t_b$, forming an edge.
Let us revisit the graph in Figure \ref{fig:method}. In this graph, we include the tools ``\emph{Validate}'', ``\emph{Login}'', and ``\emph{UpdateEmail}'' as separate nodes,
and construct two edges linking the tool nodes: ``\emph{Login}'' to ``\emph{Validate}'', ``\emph{UpdateEmail}'' to ``\emph{Login}'', respectively.

Then, based on the tool dependency graph, we adopt graph convolution \cite{kipf2016semi} to learn tool representations, where the tool dependency information is fully incorporated. Formally, we follow \citet{kipf2016semi} to conduct graph-based tool encoding in the following way:
\begin{equation}
    G(X, A)=D^{-\frac{1}{2}}(A+I)D^{-\frac{1}{2}}X.
    \label{eq:layer}
\end{equation}
Here $X$ stands for the tool embedding matrix. $A$ and $D$ denote the adjacency matrix and degree matrix of the graph respectively. There are several ways to initialize the tool embeddings here. For latter experiment, we follow \citet{qin2023toolllm} to use the retriever to encode the tool documents with a specific format for ToolBench to embeddings. While for API-Bank, we only encode the tool descriptions to mitigate the difference between the query domain and tool document domain. It is also worth noting that Equation \ref{eq:layer} removes the trainable parameters of GCN \cite{kipf2016semi} to accelerate the retrieval process.

\subsection{Online Retrieval}
The final process of TGR is to retrieve tools with the updated tool representations, which have incorporated the dependency information. Specifically, given a user query, we encode the query to an embedding vector with the same dimension as the updated tool representations. Following \citet{qin2023toolllm}, we compute the similarities between the embeddings of queries and tools as the retrieval score. Subsequently, we rank all the candidate tools in descending order and return top-$k$ tools with the highest scores.

%% file: 4-experiment.tex
\section{Experiment}

In this section, we conduct comprehensive experiments and in-depth analyses to evaluate the effectiveness of TGR.

\subsection{Setup}

\paragraph{Datasets}

We carry out experiments on two commonly-used datasets:

\begin{itemize}
    \item \textbf{API-Bank} \cite{li2023api}. The test dataset of API-Bank involves 3 levels, including a total of 311 test samples which are composed of the user query, the corresponding tools, and the final execution results. During the evaluation, we extract user queries and the corresponding tool retrieval results to quantify the tool retrieval performance.
    \item \textbf{ToolBench} \cite{qin2023toolllm}. 
    Considering the massive number of APIs and time complexity, we conduct experiments with the ToolBench instances at the I1 level\footnote{At I2 and I3 levels, each query involves APIs across different categories, which proposes challenges of high time complexity for constructing graphs. Thus, we leave extending our retrieval to other levels as future work.}.
    Given the category information about the APIs in ToolBench, we first group these APIs based on their categories. Subsequently, we identify the dependencies between APIs within each group and build a graph.
    Finally, on the basis of the graph, all API representations are updated for retrieval. 
\end{itemize}

\paragraph{Baselines}
We compare TGR with several commonly used retrieval baselines, which can be mainly divided into the following two categories:
\begin{itemize}
    \item \textbf{Word frequency-based retrieval methods}. Typically, these methods compute the similarities between the queries and tool descriptions according to the word frequency. In this category, the commonly used methods include \textbf{BM25} \cite{robertson2009probabilistic} and \textbf{TF-IDF} \cite{ramos2003using}.
    \item \textbf{Text embedding-based retrieval methods}. The methods we consider in this category involve different text embedding models: \textbf{Paraphrase MiniLM-L3-v2} \cite{reimers2019sentence} and \textbf{ToolBench-IR} \cite{qin2023toolllm}, which have been used in API-Bank and ToolBench as tool retrievers respectively.
\end{itemize}

\renewcommand{\arraystretch}{1.3}
\begin{table*}[t]
\centering
\scalebox{0.9}{
\begin{tabular}{l|l|cc|cc|cc}
\toprule
\multirow{2}{*}{\centering \textbf{Dataset}} & \multirow{2}{*}{\centering \textbf{Method}} & \multicolumn{2}{c|}{\textbf{Recall}} & \multicolumn{2}{c|}{\textbf{NDCG}} & \multicolumn{2}{c}{\textbf{Pass Rate}} \\ 
\cline{3-8}
 &  & @5 & @10 & @5 & @10 & @5 & @10 \\ 
\midrule
\multirow{6}{*}{API-Bank} 
& BM25 \citep{robertson2009probabilistic} & 0.391 & 0.493 & 0.353 & 0.394 & 0.228 & 0.302 \\ \cline{2-8}
& TF-IDF \citep{ramos2003using} & 0.566 & 0.746 & 0.501 & 0.573 & 0.383 & 0.605 \\ \cline{2-8}
& PMLM-L3-v2 \citep{reimers2019sentence} & 0.659 & 0.763 & 0.569 & 0.609 & 0.479 & 0.592 \\ \cline{2-8}
& PMLM-L3-v2+TGR & \textbf{0.736} & \textbf{0.834} & \textbf{0.622} & \textbf{0.659} & \textbf{0.576} & \textbf{0.698} \\ \cline{2-8}
& ToolBench-IR \citep{qin2023toolllm} & 0.714 & 0.790 & 0.639 & 0.670 & 0.531 & 0.624 \\ \cline{2-8}
& ToolBench-IR+TGR & \textbf{0.761} & \textbf{0.878} & \textbf{0.664} & \textbf{0.712} & \textbf{0.595} & \textbf{0.788} \\ 
\midrule
\multirow{6}{*}{ToolBench-I1} 
& BM25 \citep{robertson2009probabilistic} & 0.175 & 0.218 & 0.224 & 0.221 & 0.030 & 0.090 \\ \cline{2-8}
& TF-IDF \citep{ramos2003using} & 0.406 & 0.525 & 0.442 & 0.473 & 0.210 & 0.330 \\ \cline{2-8}
& PMLM-L3-v2 \citep{reimers2019sentence} & 0.365 & 0.468 & 0.399 & 0.421 & 0.140 & 0.250 \\ \cline{2-8}
& PMLM-L3-v2+TGR & \textbf{0.429} & \textbf{0.556} & \textbf{0.451} & \textbf{0.483} & \textbf{0.240} & \textbf{0.450} \\ \cline{2-8}
& ToolBench-IR \citep{qin2023toolllm} & 0.709 & 0.841 & 0.791 & 0.807 & 0.460 & 0.690 \\ \cline{2-8}
& ToolBench-IR+TGR & \textbf{0.742} & \textbf{0.868} & \textbf{0.811} & \textbf{0.829} & \textbf{0.510} & \textbf{0.730} \\ 
\bottomrule
\end{tabular}
}
\caption{Evaluation results on API-Bank and ToolBench-I1.}
\label{tab:main}
\end{table*}

\begin{table}[ht]
\centering
\begin{tabular}{l|c|c}
\toprule
  & \textbf{Valid} & \textbf{Test} \\
\midrule
  Precison & 0.775 & 0.893 \\
  Recall & 0.814 & 0.760 \\
  F1 & 0.792 & 0.817 \\
\bottomrule
\end{tabular}
\caption{Performance of the tool dependency discriminator. We evaluate the Precision, Recall, and F1 score on the train, valid, and test datasets.}
\label{tab:discriminator}
\end{table}

\paragraph{Implementation Details}
We use BERT-base-uncased \cite{devlin2018bert} as the base model of the discriminator. As described in Section \ref{sec:discriminator}, we first pretrain the discriminator on the category-balanced pretraining dataset, and then finetune it on the category-imbalanced finetuning dataset. During this process, we keep the checkpoint with the best performance on the validation dataset and evaluate its performance on the test dataset. Finally, we use Precision, Recall, and F1 score as the evaluation metrics for the discriminator.

As for the tool retrieval experiment on API-Bank, we simply use the description of tools for retrieval. For ToolBench, we follow \citet{qin2023toolllm} to use a structured document format of tools containing names, descriptions, and parameters for retrieval. Lastly, following \citet{qu2024colt}, we consider three metrics: Recall, NDCG, and Pass Rate at the settings of top-5 and top-10 for both API-Bank and ToolBench. 
Here we define the Pass Rate as the proportion of test samples whose required tools are totally retrieved successfully, which can be formalized as follows:
\begin{equation}
    pass@k=\frac{1}{|Q|}\sum_{q}^{Q}\mathbb{I}(\Phi(q) \subseteq \Psi^k(q))
\end{equation}
where $\Phi(q)$ denotes the set of ground-truth tools for query $q$, $\Psi^k(q)$ represents the top-$k$ tools retrieved for query $q$, and $\mathbb{I}(\cdot)$ is an indicator function that returns 1 if the retrieval results include all ground-truth tools within the top-$k$ results for query $q$, and 0 otherwise.

A higher Recall demonstrates that more required tools are successfully retrieved, a higher NDCG score indicates that the target tools achieve higher ranks, and a higher Rass Rate signifies that more queries are completed with all the required tools retrieved.

\subsection{Discriminator Performance}

\label{sec:exp_discriminator}

\begin{table}
\centering
\begin{tabular}{l|c|c}
\toprule
  & \textbf{API-Bank} & \textbf{ToolBench} \\
\midrule
\#Total  & 119 & 10,439 \\
\#Connected & 50 & 8,600 \\ \cline{1-3}
Proportion & 0.420 & 0.824 \\
\bottomrule
\end{tabular}
\caption{The proportion of connected graph nodes in API-Bank and ToolBench.}
\label{tab:nodes}
\end{table}

In this group of experiments, we first focus on the quality of the constructed tool dependency graph, which, intuitively, greatly depends on the discriminator and is crucial for the performance of TGR.

To this end, we present the Precision, Recall, and F1 score of our discriminator across the validation and testing datasets in Table \ref{tab:discriminator}. Overall, our discriminator can achieve decent performance on two datasets. Additionally, the resulting graphs are visualized in Appendix \ref{sec:vis}.

Furthermore, we calculate the proportions of connected nodes in the tool dependency graphs, as shown in Table \ref{tab:nodes}.
We note that the proportions of connected graph nodes differ between the two datasets, which is influenced by the granularity of tool functions because fully-featured tools are less likely to depend on others while specialized tools designed with fine-grained functions usually have more intensive dependencies.

\begin{table*}[]
    \centering
    \begin{tabular}{l|l|cc|cc|cc}
    \toprule
    \multicolumn{2}{c|}{\multirow{2}{*}{\textbf{Method}}}
     & \multicolumn{2}{c|}{\textbf{Recall}} & \multicolumn{2}{c|}{\textbf{NDCG}} & \multicolumn{2}{c}{\textbf{Pass Rate}} \\ \cmidrule(lr){3-4} \cmidrule(lr){5-6} \cmidrule(lr){7-8}
     \multicolumn{2}{c|}{} & @5 & @10 & @5 & @10 & @5 & @10 \\ 
     \midrule
     \multirow{2}{*}{\makecell[l]{PMLM-L3-v2 \\ \cite{reimers2019sentence}}}
     & +TGR-d  & 0.736 & 0.834 & 0.622 & 0.659 & 0.576 & 0.698 \\
     & +TGR-m & \textbf{0.745} & \textbf{0.846} & \textbf{0.634} & \textbf{0.672} & \textbf{0.592} & \textbf{0.711} \\ \hline
     \multirow{2}{*}{\makecell[l]{ToolBench-IR \\ \cite{qin2023toolllm}}}
     & +TGR-d & 0.761 & 0.878 & 0.664 & 0.712 & 0.595 & 0.788 \\
     & +TGR-m & \textbf{0.788} & \textbf{0.893} & \textbf{0.698} & \textbf{0.741} & \textbf{0.646} & \textbf{0.817} \\
    \bottomrule
    \end{tabular}
    \caption{Performance comparison between different TGRs, of which tool dependency graphs are constructed by our discriminator (represented as +TGR-d) and manual annotations (represented as +TGR-m). These group of experiments are conducted on the API-Bank \cite{li2023api}.}
    \label{tab:manual}
\end{table*}

\subsection{Main Results}
\label{sec:main}

The results of tool retrieval are presented in Table \ref{tab:main}, showing that on all three metrics, TGR significantly improves the performance of base text embedding models and outperforms word frequency-based retrieval methods to a large extent. This indicates that incorporating tool dependency as additional information greatly enhances the effectiveness of tool retrieval. Furthermore, we arrive at the following interesting conclusions.

Firstly, when applying TGR to ToolBench-IR, which is finetuned specifically for the tool retrieval task, it can achieve the SOTA performance on both datasets. Therefore, we believe that finetuning and TGR are two methods that are compatible with each other and thus can be used to improve the performance of tool retrieval simultaneously.

It can also be seen that the methods based on ToolBench-IR greatly surpass others on ToolBench. 
This is because the tool documents in ToolBench have a specific format that only ToolBench-IR can fit well since it is finetuned on the training set of ToolBench. 

\subsection{Effect of Different Dependency Graph}

In this subsection, we study the effect of the graph construction quality for TGR. 
Due to the extensive number of tools in ToolBench, which makes manual annotation of the entire graph impractical, we choose API-Bank and the same two text embedding models: Paraphrase MiniLM-L3-v2 \cite{reimers2019sentence} and ToolBench-IR \cite{qin2023toolllm} for this experiment. We also use the same metric as the main experiments in Section~\ref{sec:main}. 

Table \ref{tab:manual} lists the experimental results.
To avoid confusion, we term the TGR based on the discriminator as +TGR-d and on manual annotation as +TGR-m. From this table, we can clearly observe that 
+TGR-m performs better than +TGR-d, no matter which embedding model is used. In our opinion, this result is reasonable because the quality of the manually-constructed tool dependency graph is higher than that of the discriminator-constructed graph.  
Thus, we believe that how to improve the performance of our discriminator is very important for the further improvement of TGR. 

\subsection{Effect of Graph Density}

\begin{figure}
    \centering
    \includegraphics[width=\linewidth]{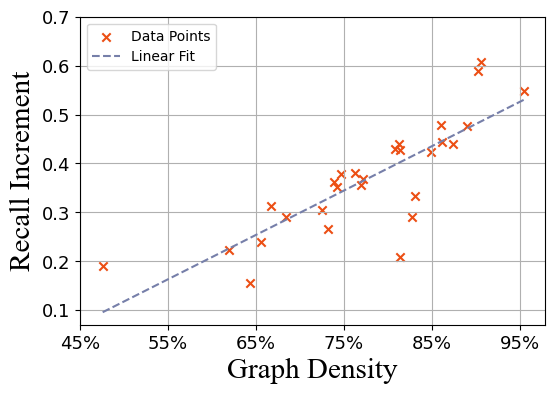}
    \caption{The relationship between the density of the tool dependency graph and the recall increment.}
    \label{fig:density}
\end{figure}

In this subsection, we evaluate the effect of the density of the tool dependency graph on tool retrieval. 
Specifically, we collate all the tools in ToolBench by their categories and rank the categories according to their graph density, which is measured by the proportion of connected tool nodes. 
Due to the limited size of the test set, we extract 100 queries for each category from the train set for evaluation, which are completely unused during the procedure of discriminator dataset construction. 
For the evaluation metric, we measure the recall increment of the TGR-enhanced text embedding model over the base text embedding model at the top-5 setting. Here we use the ToolBench-IR as the text embedding model considering its excellent retrieval performance.

The result is shown in Figure~\ref{fig:density}. We can see that as the density of the graph increases, the recall increment also exhibits an upward trend, which validates that dependencies between tools indeed help to improve the performance of tool retrieval. It also demonstrates that TGR is highly robust and more effective for dependency-intensive tool retrieval.

\subsection{Case Study}

\begin{table}[]
\centering
\scalebox{0.8}{
    \renewcommand{\arraystretch}{1.3}
    \begin{tabular}{m{3.5cm}|m{5.5cm}}
    \toprule
        Query & Can you please help me delete my account? My username is foo and my password is bar. \\ 
        \midrule
        Ground Truth & \textcolor{blue}{GetUserToken} \newline \textcolor{blue}{DeleteAccount} \\
        \midrule
        Dependency & DeleteAccount $\to$ GetUserToken \\
        \midrule
        \midrule
        ToolBench-IR & 1. \textcolor{blue}{DeleteAccount} \newline 2. AccountInfo \newline 3. DeleteReminder \newline 4. DeleteBankAccount \newline 5. DeleteScene \\
        \hline
        Toolbench-IR+TGR & 1. \textcolor{blue}{GetUserToken} \newline 2. Transfer \newline 3. OpenBankAccount \newline 4. RegisterUser \newline 5. \textcolor{blue}{DeleteAccount}\\
    \hline
        
    \end{tabular}
    }
    \caption{Case study of tool retrieval on API-Bank. Correct APIs are highlighted in \textcolor{blue}{blue}.}
    \label{tab:case_apibank}
\end{table}

\begin{table}[]
    \centering
    \scalebox{0.8}{
    \renewcommand{\arraystretch}{1.3}
    \begin{tabular}{m{3.5cm}|m{6cm}}
    \toprule
        Query & Which football leagues' predictions are available for today? I want to explore the predictions for the Premier League and La Liga. \\
        \midrule
        Ground Truth & \textcolor{blue}{Get Today's Predictions} \newline \textcolor{blue}{Get Next Predictions} \\
        \midrule
        Dependency & Get Next Predictions $\to$ Get Today's Predictions \\
        \midrule
        \midrule
        ToolBench-IR & 1. Daily Predictions \newline 2. Football predictions by day \newline 3. \textcolor{blue}{Get Next Predictions} \newline 4. VIP Scores \newline 5. Prediction DetPredictionails \\
        \hline
        ToolBench-IR+TGR & 1. Football predictions by day \newline 2. Basketball predictions by day \newline 3. \textcolor{blue}{Get Today's Predictions} \newline 4. \textcolor{blue}{Get Next Predictions} \newline 5. Sample predictions \\
    \hline
    \end{tabular}
    }
    \caption{Case study of tool retrieval on ToolBench. Correct APIs are highlighted in \textcolor{blue}{blue}.}
    \label{tab:case_toolbench}
\end{table}

Finally, we provide two examples to further illustrate how TGR improves the performance of tool retrieval. We conduct case studies on both API-Bank and ToolBench with ToolBench-IR as the base text embedding model, since it achieves the best performance in our main experiments.

Table~\ref{tab:case_apibank} presents the first example in API-Bank, where the tool ``\emph{DeleteAccount}'' requires the result (the user token) from the tool ``\emph{GetUserToken}'' as an input parameter. We display the retrieval results by their ranking orders. The retrieval results of ToolBench-IR contain only one correct API ``\emph{DeleteAccount}'' with the top rank due to its high semantic similarity with the query. With the enhancement of TGR, ``\emph{GetUserToken}'', which ``\emph{DeleteAccount}'' depends on, incorporates the information from ``\emph{DeleteAccount}'' and is also  retrieved with a high rank.

Table~\ref{tab:case_toolbench} presents the second example in ToolBench. It is obvious that the base ToolBench-IR misses the required API ``\emph{Get Today's Prediction}''. Given the relationship that ``\emph{Get Next Prediction}'' depends on ``\emph{Get Today's Prediction}'', the TGR-enhanced ToolBench-IR succeeds in retrieving the missing tool.